# Comments on the paper: Structural, optical properties and effect of amino acid on growth of KDP crystals

Bikshandarkoil R Srinivasan[a] & K R Priolkar[b]
[a]Department of Chemistry, Goa University, Goa 403206, INDIA Email: srini@unigoa.ac.in
[b]Department of Physics, Goa University, Goa 403206, INDIA Email: krp@unigoa.ac.in

**Abstract**

The *L*-arginine mixed potassium dihydrogenphosphate (KDP) crystal reported by Saravanan et al *Indian J Pure App Phys* 51 (2013) 254, is a dubious crystal. The reported unit cell parameters contradict the definition of the tetragonal crystal system.

**Keywords**: tetragonal crystal system; potassium dihydrogenphosphate; *L*-arginine; dubious crystal.

**Comment**

Phosphate salts having the general formula $MH_2PO_4$, are of technological interest because they exhibit ferroelectric (M = K, Rb, Cs) or antiferroelectric (M = $NH_4$) properties at low temperatures[1]. Of the alkali metal phosphates, potassium dihydrogen phosphate (KDP) is a well-known solid and is used as a reference material to report second harmonic generation (SHG) efficiency in nonlinear optical (NLO) materials research[2]. The dihydrogen phosphates of $K^+$ (KDP)[3] and ammonium (ADP)[4] are isomorphous at room temperature and crystallize in the tetragonal system in the non-centrosymmetric space group $I\bar{4}2d$ (Table 1).

Table 1.   Unit cell parameters of the isostructural KDP and ADP

| Compound | $a$ (Å) | $b$ (Å) | $c$ (Å) | α = β = γ (°) | V (Å$^3$) | Reference |
|---|---|---|---|---|---|---|
| ADP | 7.502 | 7.502 | 7.546 | 90.0 | 424.69 | 4 |
| KDP | 7.452 | 7.452 | 6.974 | 90.0 | 387.28 | 3 |
| L-arginine doped KDP | **6.307** | **10.536** | **6.299** | --- | 385.6[#] | Title paper |

[#]The correct volume for L-arginine doped KDP should be 418.57 Å$^3$

KDP is a well-known solid and has been the subject of several research investigations. In an earlier study Xu and Xue[5] have investigated the growth of KDP crystals in an aqueous solution by the addition of ethanol ($CH_3$-$CH_2$-OH). These workers have reported that the crystal structure of the as-grown crystallites remains practically unchanged, demonstrating that ethanol molecules could not enter the lattice sites (did not get doped / incorporated) to change the crystal structure. The only change that occurred was in the morphology of the KDP crystallites with the shape becoming slender[5]. In view of this work by Xu and Xue[5] the paper entitled *"Structural, optical properties and effect of amino acid on growth of KDP crystals"* by Saravanan et al[6] reporting mixing of *L*-arginine into the KDP lattice attracted our attention. We believe that the authors of this paper use the term 'mixing' by which they actually mean 'doping'. However, the claim of incorpoarting *L*-arginine which is a larger molecule than ethanol (Fig. 1) into the KDP lattice is not only unsubstantiated but also incorrect as will be proved below.

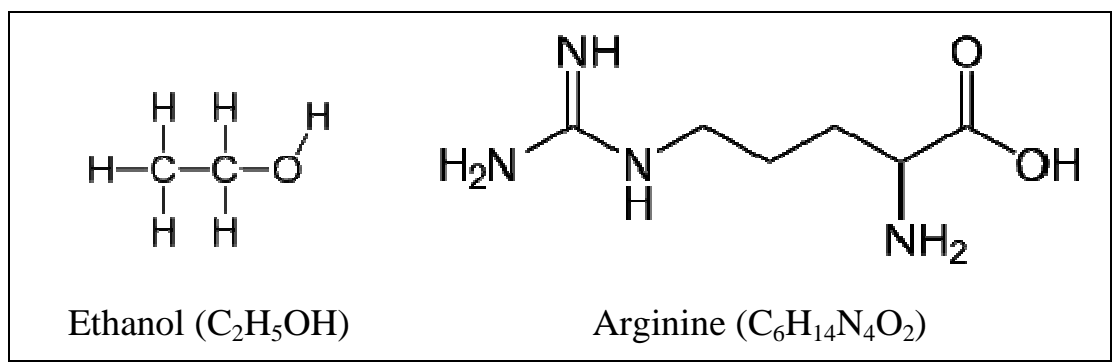

Ethanol ($C_2H_5OH$)   Arginine ($C_6H_{14}N_4O_2$)

Fig. 1 Structure of ethanol and arginine (Note: C atom bonded to the –COOH is a chiral carbon)

The authors of the title paper report to have grown crystals of *L*-arginine mixed (doped) KDP, by addition of 5 weight % of *L*-arginine into a saturated solution of KDP[6]. Based on a study of the X-ray powder pattern the authors reported that the grown crystal belongs to the tetragonal system having lattice parameters $a$ = 6.307 Å, $b$ = 10.536 Å and $c$ = 6.299 Å.

The tetragonal crystal system has a four-fold rotation axis as an essential symmetry element and its unit cell parameter *a* equals *b*. The relation $a = b \neq c$ ; $\alpha = \beta = \gamma = 90°$, characterizes a tetragonal crystal system and is described in all standard text books[7]. When $a \neq b \neq c$ and $\alpha = \beta = \gamma = 90°$, the crystal system is orthorhombic, which is devoid of the fourfold axis. The reported unit cell ($a = 6.307$ Å, $b = 10.536$ Å and $c = 6.299$ Å) in the title paper appears to redefine a tetragonal crystal and such a classification with unequal values for *a, b* and *c* not only contradicts all known laws of crystal physics but is also not in accordance with the definition of a tetragonal crystal system in the International Tables for Crystallography[8]. Since $6.307 \neq 10.536 \neq 6.299$ Å, the reported data in the title paper can at best be described as an orthorhombic cell (if $\alpha = \beta = \gamma = 90°$) and never a tetragonal cell.

In order to justify, that such a crystal system is tetragonal the authors reported the unit cell volume for their so called tetragonal cell as 385.60 Å$^3$, a value close to but different from that of the cell volume (387.28 Å$^3$) of KDP. But the authors chose a wrong value, because the cell volume (6.307 × 10.536 × 6.299) is actually 418.57 Å$^3$. The reporting of unequal values for *a, b* and *c* has been done to show the incorporation or the so called mixing of *L*-arginine into the KDP lattice as can be evidenced from the statement of the authors in the title paper, *'The crystal was identified by comparing the interplanar spacing and the intensities of the powder pattern with the JCPDS data of KDP crystal. The calculated lattice parameters are found to change as seen from Table 1. Mixing changes the cell axes and hence the cell volume'*. Thus the authors incorrectly claim that the mixing (incorporation / doping) of *L*-arginine has resulted in an exchange of *b* and *c* axes, which does not make any crystallography sense. The questionable nature of these results can be further evidenced by the indexing of the reported powder pattern (Fig. 2) wherein the (0 0 2) reflection is shown at $2\theta = 58°$ while the (2 2 0) or (5 3 0) or (6 3 2) reflections are assigned at $2\theta$ values < 58° which is just not possible. A comparison of the powder pattern in the title paper with that of an authentic pattern (Fig. 3) of KDP (JCPDS data) not only reveals that all Miller indices are incorrect, but also indicates additional spurious and unindexed reflections indicating a mixture of phases. Thus the title paper reports an inappropriate crystal symmetry, incorrect assignment of Miller indices and unit cell data incompatible with the powder pattern. Hence, the characterization of *L*-arginine mixed KDP crystal is improper and unacceptable.

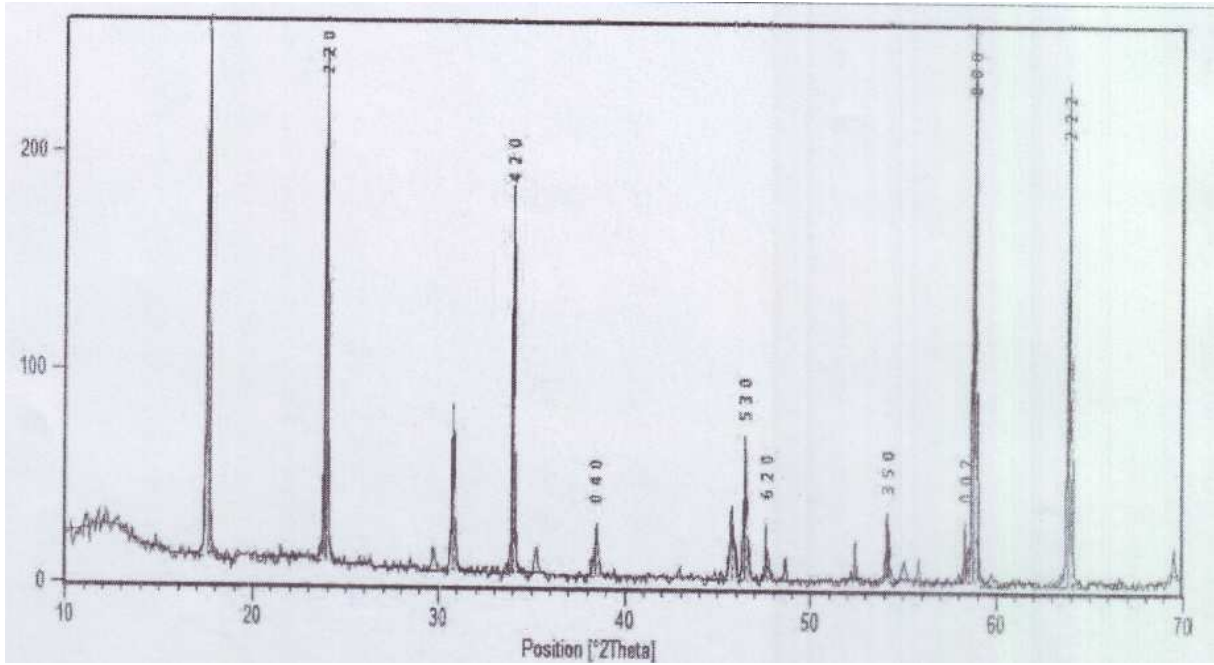

Fig. 2 Reported powder pattern reported for a so called *L*-arginine mixed potassium dihydrogen phosphate (KDP) crystal. Taken from the title paper Ref. 6. *All Miller indices in the pattern are wrong*. For an authentic powder pattern of KDP pl. see Fig. 3

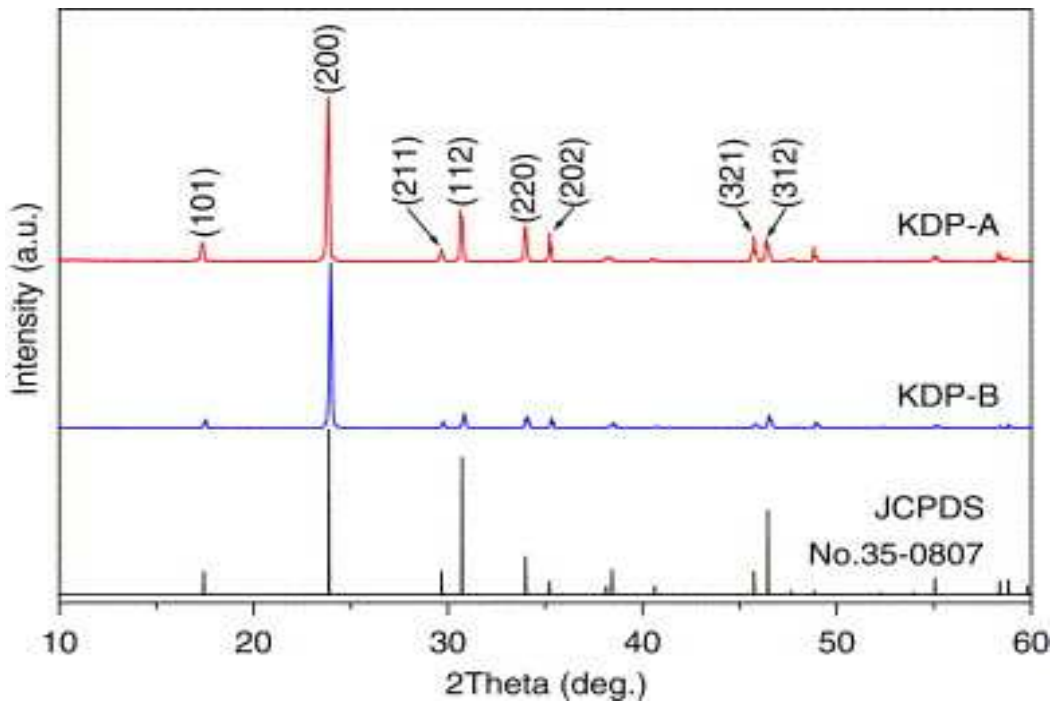

Fig. 3 Powder pattern reported for KDP crystallites grown from aqueous solution (KDP-A), KDP crystallites grown from aqueous solution in the presence of ethanol (KDP-B). The powder pattern of KDP from JCPDS is also given. **Taken from Ref. 5** by Xu D & Xue D, *J Cryst Growth* 286 (2006) 108.